\documentclass[aps,prd,twocolumn,superscriptaddress,amsfont,graphicx,nofootinbib,preprintnumbers]{revtex4-1}%

\UseRawInputEncoding
\usepackage{color,graphicx,epsfig}
\usepackage{ifpdf}
\usepackage{amsmath}
\usepackage{bm}
\usepackage[english]{babel}
\usepackage{amssymb}
\usepackage{braket}
\usepackage{hyperref}
\usepackage{enumerate}
\usepackage{url}
\usepackage{multirow}
\usepackage{threeparttable}
\usepackage{makecell}
\usepackage{setspace}

\usepackage{slashed}

\usepackage{changes}

\begin{document}

\title{Polarization Signals from Axion-Photon Resonant Conversion in Neutron Star Magnetosphere}%

% Force line breaks with \\
%%\thanks{A footnote to the article title}%

\author{Ningqiang Song}
\email{songnq@itp.ac.cn}
\affiliation{Institute of Theoretical Physics, Chinese Academy of Sciences, Beijing, 100190, China}

\author{Liangliang Su}
\email{liangliangsu@njnu.edu.cn}
\affiliation{Department of Physics and Institute of Theoretical Physics, Nanjing Normal University, Nanjing, 210023, China}

\author{Lei Wu}
\email{leiwu@njnu.edu.cn}
\affiliation{Department of Physics and Institute of Theoretical Physics, Nanjing Normal University, Nanjing, 210023, China}

\date{\today}% It is always \today, today,

\begin{abstract}

Neutron stars provide ideal astrophysical laboratories for probing new physics beyond the Standard Model. If axions exist, photons can develop linear polarization during photon-axion conversion in the magnetic field of a neutron star. We find that the plasma in the neutron star magnetosphere could dramatically enhance the polarization through the resonant conversion effect. With the polarization measurements from PSR B0656+14, 4U 0142+61, and the benchmark polarization measurement in the mid-infrared band, we demonstrate that optical and infrared polarization from neutron stars can provide strong constraints on the axion-photon coupling over a broad axion mass range $10^{-11}\lesssim m_a \lesssim 10^{-3}$ eV.

\end{abstract}

\maketitle

%\tableofcontents

\section{Introduction}
Axions and axion-like particles are one of the prominent candidates in the extension of the Standard Model (SM)~\footnote{For simplicity, we use the term ``axions" interchangeably for both QCD axions and axion-like particles, allowing the axion mass and axion-photon coupling to vary independently.}. In particular, QCD axion offers an intriguing solution to the strong \textit{CP}-problem~\cite{Peccei:1977hh,Peccei:1977ur,Weinberg:1977ma,Kim:1979if} in particle physics. Axions may also arise from string theory compactifications and comprise cosmological dark matter through a variety of mechanisms~\cite{Preskill:1982cy, Abbott:1982af, Dine:1982ah, Duffy:2009ig,Athron:2020maw,Semertzidis:2021rxs,Chadha-Day:2021szb,Gu:2021lni,Adams:2022pbo,Gong:2023ilg}.

The wide range of axion mass and feeble interactions between axions and SM particles pose challenges in axion searches. Current axion experiments rely on the electromagnetic coupling of axions in the form of $\mathcal{L}\supset -\frac{1}{4}g_{a\gamma}a\mathbf{E}\cdot\mathbf{B}$, where axions and photon inter-convert in the presence of the (strong) magnetic field. These include light shining through walls~\cite{Ehret:2010mh,Redondo:2010dp,Bahre:2013ywa}, axion helioscopes (e.g. CAST~\cite{CAST:2007jps,CAST:2017uph}), axion haloscopes through axion-photon resonant conversion in cavities (e.g. ADMX~\cite{ADMX:2009iij,ADMX:2018gho,ADMX:2019uok,ADMX:2021nhd,ADMX:2018ogs,Crisosto:2019fcj}, gamma ray searches~\cite{Fermi-LAT:2016nkz,Meyer:2016wrm,Meyer:2020vzy,Davies:2022wvj} through axion-photon oscillations, among others~\cite{Guo:2023hyp,Arza:2023rcs}.

On the astrophysical side, the exclusive plasma environment in the atmosphere of (compact) stars facilitates the resonant conversion between axion~\cite{Hook:2018iia,Huang:2018lxq,Safdi:2018oeu,Buckley:2020fmh,Witte:2021arp,Millar:2021gzs,Battye:2021xvt,Battye:2021yue,Wang:2021hfb,Foster:2022fxn,Battye:2021xvt,Zhou:2022yxp,Chattopadhyay:2023nuq} or dark photon dark matter~\cite{An:2022hhb,Hardy:2022ufh,An:2023wij,An:2023mvf} and photons, producing spectral-line type features that attract extensive astrophysical searches. However, such conversion relies on the resemblance between the plasma frequency and axion mass, which applies merely to a limited mass range with axion as dark matter as a prerequisite. 
On the other hand, the polarization signal from compact stars may open a new avenue for searching for axions. In the presence of the axion-photon coupling, the photon component aligned with the magnetic field could convert to axions, while the orthogonal photon component remains intact. Consequently, the photon from stars will acquire a linear polarization after traversing the star magnetic field~\cite{Jain:2002vx,Lai:2006af,Gill:2011yp,Payez:2011sh,Perna:2012wn, PhysRevD.96.043519,Galanti:2022iwb,Dessert:2022yqq,Fortin:2023jlg,Gan:2023swl,Gau:2023rct}. Although a strong magnetic field is required to boost the axion-photon mixing, it also tends to suppress the conversion probability due to the vacuum polarization effect inspired by the Euler-Heisenberg interaction.

In this work, we investigate the axion-induced polarization signal from neutron stars. In contrast with the cases in magnetic white dwarfs (MWDs)~\cite{Gill:2011yp,Dessert:2022yqq}, we consider orders of magnitude strong magnetic fields around neutron stars. In particular, we incorporate the resonant conversion effect to bypass the suppression from vacuum polarization, which is much more efficient than the non-resonant conversion process adopted in previous works. This is achieved by including the plasma effect which is usually omitted in similar studies. Also distinct from the spectral-line search, the resonant conversion may take place when the plasma frequency is close to the vacuum contribution, and hence the observation of the polarization at a single frequency allows to place constraints on a large range of axion mass, without relying on axion as dark matter concomitantly. As a proof of concept, we derive the sensitivity on axion-photon coupling by analyzing the polarization measurements of neutron star  PSR B0656+14 and 4U 0142+61, as well as a benchmark polarization measurement in the infrared band, which are proven to be stronger than the current constraints for axion with mass $10^{-11} \; \mathrm{eV}  \lesssim m_a \lesssim 10^{-3}$ eV.

\section{photon-axion Resonant Conversion}
\label{sec:resonantconversion}
We first consider the evolution of the axion-photon system in the neutron star magnetosphere with a magnetic field up to $10^{15}$~G, which is described by the Lagrangian~\cite{PhysRevD.37.1237},
\begin{equation}
\begin{aligned}
\mathcal{L}= & -\frac{1}{4} F_{\mu \nu} F^{\mu \nu}+\frac{1}{2}\left(\partial_{\mu} a \partial^{\mu} a-m_{a}^{2} a^{2}\right)-\frac{g_{a \gamma}}{4 } F_{\mu \nu} \tilde{F}^{\mu \nu} a \\
& +\frac{\alpha^{2}}{90 m_{e}^{4}}\left[\left(F_{\mu \nu} F^{\mu \nu}\right)^{2}+\frac{7}{4}\left(F_{\mu \nu} \tilde{F}^{\mu \nu}\right)^{2}\right],
\end{aligned}
\end{equation}
where $m_a$ is the mass of the axion filed $a$ and  $g_{a \gamma}$ is the axion-photon mixing. The second line originates from the non-linear vacuum polarization effect in the presence of strong magnetic field. For a relativistic axion and photon with energy $\omega$, this leads to the equation of motion
\begin{equation}
\left[i \partial_z + \left(\begin{array}{ccc}
\Delta_{a}  & \Delta_{\mathrm{M}} & 0 \\
\Delta_{\mathrm{M}}  &\Delta_{\|}+\Delta_{\mathrm{pl}} & 0 \\
0 & 0 & \Delta_{\perp} 
\end{array}\right)\right] \left(\begin{array}{c}
a \\
A_{\|} \\
A_{\perp}
\end{array}\right)=0,
\label{eq:motion}
\end{equation}
where 
\begin{equation}
\begin{aligned}
\Delta_{\mathrm{M}} & =\frac{1}{2} g_{a \gamma} B \sin \Theta,\;\; \Delta_{a}  = - \frac{m_{a}^{2}}{2 \omega}, \;\; \Delta_{\mathrm{pl}}  = -\frac{\omega_{\mathrm{pl}}^{2}}{2 \omega}, \\
\Delta_{\|} & =\frac{7}{2} \omega \xi \sin^{2} \Theta, \;\; \Delta_{\perp} =\frac{4}{2} \omega \xi \sin^{2} \Theta, 
\end{aligned}
\end{equation}
with $\xi = (\alpha/45 \pi) (B/B_{\mathrm{QED}})^2$, the critical QED magnetic field strength $B_{\mathrm{QED}} = m_e^2/e \approx 4.41 \times 10^{13} \; \mathrm{G}$, and the angle $\Theta$ between the external magnetic field and the direction of photon's propagation. $\Delta_{\mathrm{pl}}$ describes the plasma effect arising from the modification of the photon dispersion relation in finite charge density. As the magnetosphere is dominated by electrons, the plasma frequency is related to the free-electron density by $\omega_{\mathrm{pl}}  = \sqrt{4 \pi \alpha n_e /m_e}$.

It is straightforward to see that the photon polarization parallel to the plane of propagation and magnetic field, $A_{\|}$, converts to axion, while the polarization perpendicular to this plane $A_{\perp}$ does not, resulting in net linear polarization that will be addressed in the next section. The mixing between axion and photon depends on the interplay of the vacuum polarization, plasma effect, and the axion mass, and is directly proportional to the strength of the magnetic field (see below for complications). In approximately constant magnetic field and plasma, the effective axion-photon mixing angle $\beta_m$ can be obtained from the matrix diagonalization in Eq.~\eqref{eq:motion}, and the strength of mixing
\begin{equation}
\tan 2 \beta_m =\frac{2 \Delta_{\mathrm{M}}}{\Delta_{\|} + \Delta_{\mathrm{pl}} -\Delta_{a}} .
\end{equation}

Here we identify two different scenarios. MWDs are characterized by a thin atmosphere with a typical scale height below 100~m~\cite{Hardy:2022ufh}. The free-electron density diminishes exponentially so that $\Delta_{\|}\gg\Delta_\mathrm{pl}$ in the majority part of the atmosphere. Consequently, the photon-axion mixing is weak in the strong magnetic field of MWDs due to the suppression effect of vacuum polarization within the axion mass range of interest, i.e., $\tan 2 \beta_m = 2 \Delta_{\mathrm{M}}/\Delta_{\|} \propto B^{-1} \ll 1$. In this limit, the evolution of photon and axion can be obtained from the perturbative solution of the equations of motion Eq.~\eqref{eq:motion}, where the $\Delta_{\rm M}$ term is treated as perturbation (we refer to the supplementary materials for more details).

On the other hand, neutron stars and even magnetars, with an appropriate free-electron density, exhibit strong photon-axion mixing when the condition 
\begin{equation}
    \Delta_{\|} + \Delta_{\mathrm{pl}} -\Delta_{a} =0
\end{equation}
is met, leading to efficient photon-axion conversion, or resonant conversion effect. The perturbative solution will break down in the strong mixing regime, particularly in the resonance region~\cite{PhysRevD.37.1237}. Now, we turn to a more realistic neutron star magnetosphere with an inhomogeneous magnetic field. The photon-axion conversion resembles resonant neutrino oscillation in matter, which is determined by the nonadiabatic jump probability near the resonance location, and the jump probability is attained using the Landau-Zener (LZ) formula~\cite{RevModPhys.61.937} 
\begin{equation}
P_{\mathrm {jump }}=e^{-\pi \gamma_{\text {res }} / 2},
\end{equation}
with  the adiabaticity ratio
\begin{equation}
\gamma_{\mathrm{res}} = \frac{4 \Delta_M^2 H}{|\Delta_a+\Delta_{\|}|}\,,
\end{equation}
where $H = n_e/ |n_e^{\prime}|$ is the density scale height at the resonance point along the line of sight (l.o.s) (see the supplementary materials for detailed calculations). It is readily seen that the conversion probably in this scenario does not suffer from the $B^{-1}$ impression and efficient photon-axion conversion is expected in strong magnetic field~\footnote{Resonant conversion is also expected to occur in MWDs as the maximum electron density in the atmosphere is high (about $10^{17}$~cm$^{-3}$). However, the small scale height there prevents strong conversion.}.  The probability of photon-axion conversion across the resonance is expressed as $P_{\gamma \rightarrow a} = 1 - P_{\mathrm{jump}}$~\cite{Lai:2006af} and is positively correlated with the resonance adiabaticity ratio $\gamma_{\mathrm{res}}$.

We note that the LZ formula was derived based on the assumptions of constant magnetic field and linearly changing plasma density. Although these conditions are not always satisfied in realistic astrophysical environments, the formula remains effective in the non-adiabatic ($\gamma_{\mathrm{res}} < 1$) regime~\cite{Carenza:2023nck}, which is always valid in this work.

The charge particle density in the magnetosphere of a neutron star is commonly described by the Goldreich-Julian (GJ) model~\cite{goldreich1969pulsar} in the literature. The GJ model depends on the spin period $T$, the magnetic field strength at the star's surface $B_p$, and the polar angles $\theta_{\mathrm{NS}}$ and $\theta_m$,  which represent the angles of the l.o.s and the magnetic axis with respect to the rotation axis (see the supplementary materials). Despite its uncertainties, the GJ model exclusively offers a baseline for the minimal plasma density $n_{\mathrm{GJ}}$, while actual plasma densities are intricately influenced by various factors. For instance, processes such as pair creation cascades occurring at the poles have the potential to significantly augment electron density, reaching a density that is  $10^{2}$ to $10^{4}$ higher than that predicted by the GJ model~\cite{Eilek:2016hms}. Given this, we will adopt the GJ model as the benchmark and analyze situations in which plasma densities are scaled by factors of $1$, $10^{2}$, and $10^{4}$ with respect to the GJ model. 

Under the assumption of the GJ model and dipole magnetic field profile, there are two possible situations satisfying the photon-axion resonance conditions, $\Delta_a = \Delta_{\|} +\Delta_{\mathrm{pl}}$, with the following properties: 

    {\bf Plasma-vacuum resonance} $|\Delta_a|\ll \Delta_{\|} \approx - \Delta_{\mathrm{pl}}$. This resonance condition has the adiabaticity ratio
    \begin{equation}
    \gamma_{\mathrm{res}} = \frac{4 \Delta_{\mathrm{M}}^2 H}{\Delta_{\|}} \propto \frac{g_{a \gamma}^2 H}{\omega} \approx g_{a\gamma}^2 \frac{r_{\mathrm{NS}}+z_{\mathrm{res}}}{3 \omega },
    \label{eq:resvacuum}
    \end{equation}
    where $z_{\mathrm{res}}$ is the height of the resonance point from the neutron star surface. This implies that the lower energy photon is more likely to convert to axion at the higher resonance point. 

    {\bf Plasma-mass resonance} $\Delta_{\|} \ll |\Delta_{\mathrm{pl}}|\approx |\Delta_{a}|$. This situation mostly occurs far from neutron stars when the magnetic field is not strong enough for $\Delta_{\|}$ to dominate the conversion. There $m_a^2 \approx \omega_{\mathrm{pl}}^2$ and  
    \begin{equation}
        \gamma_{\mathrm{{res}}} = \frac{4 \Delta_{\mathrm{M}}^2 H }{|\Delta_a|} \propto \frac{g_{a \gamma}^2 B^2_{\mathrm{res}} \omega H }{m_a^2} \approx g_{a\gamma}^2 B_{\mathrm{res}} \omega H, 
    \end{equation}
     where $B_{\mathrm{res}}$ is the magnetic filed strength at the resonance point. It shows that stronger resonant conversion is expected at a lower resonance point and for higher energy photon. This is the case for heavier axions as $m_a \approx \omega_{\mathrm{pl}} \propto \sqrt{B}$.

\begin{figure}[ht]
  \centering
\includegraphics[width=7.5cm,height=7cm]{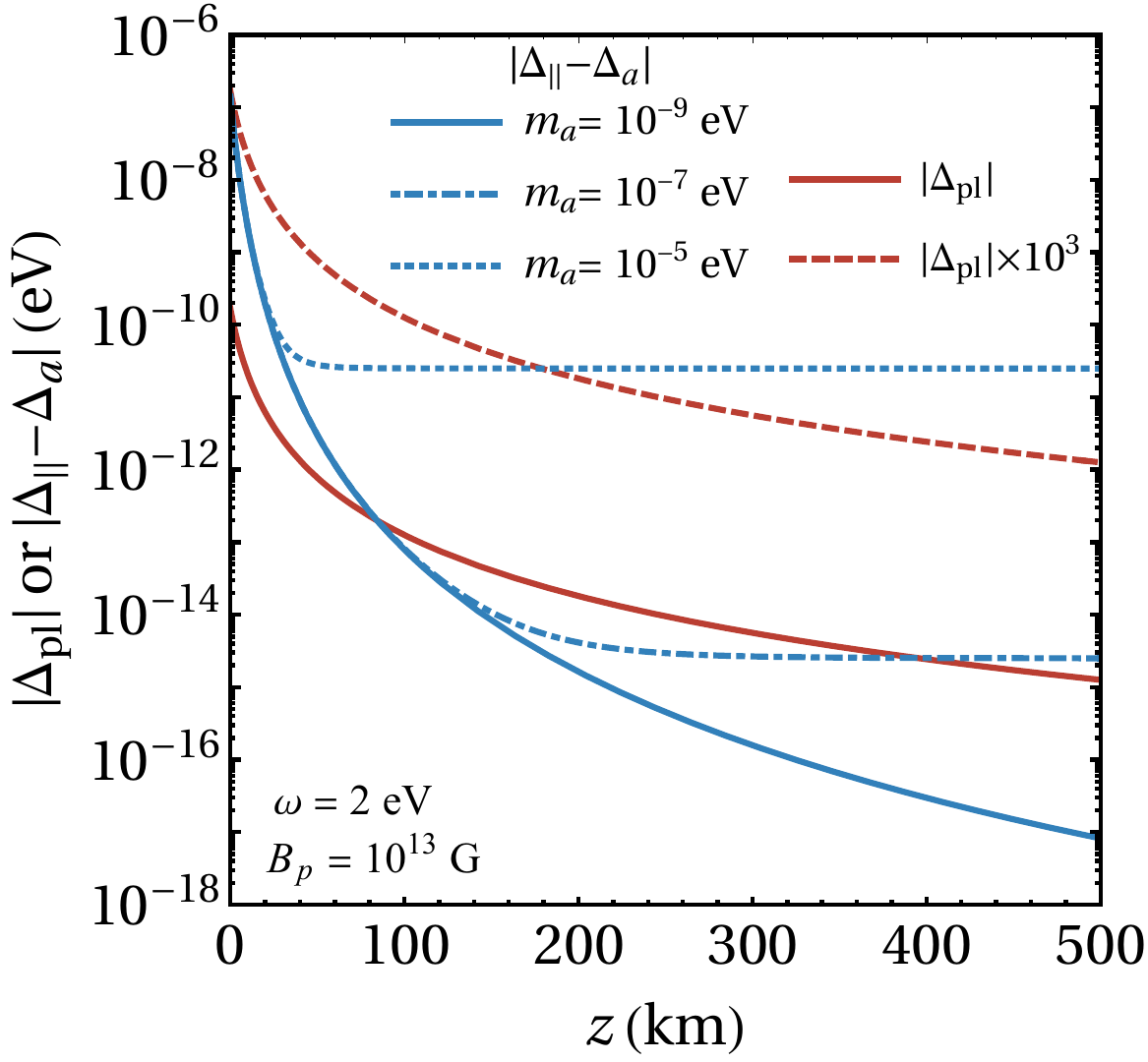}
  \caption{The $|\Delta_{\mathrm{pl}}|$ and $|\Delta_{\|}-\Delta_a|$ term as a function of the height $z$ from the neutron star surface. In which, the red solid and dashed line denote the former and its $10^{3}$ times, respectively. The blue solid, dot-dash, and dotted lines show the latter for the different axion mass $m_a = 10^{-9}$ eV, $10^{-7}$ eV, and $10^{-5}$ eV, respectively. We assume a neutron star with $B_p = 10^{13}$ G, $\omega =2 $ eV, $r_{\mathrm{NS}} = 10$ km, $T = 1$ s, $\Omega t =0^{\circ}$, $\theta_{\mathrm{NS}}= 40^{\circ}$, and  $\theta_{m}= 30^{\circ}$}
  \label{fig:resonance_example}
\end{figure}

The radial profile of the plasma and vacuum contributions are depicted in Fig.~\ref{fig:resonance_example}. In the outer magnetosphere, $|\Delta_{\|}-\Delta_a|$ is eventually dominated by $\Delta_a$. Due to the different $B$ scaling, $\Delta_{\|}$ always decreases faster than $|\Delta_{\mathrm{pl}}|$. If the axion mass is low, two types of resonances may occur subsequently when photons propagate in the magnetosphere. The resonance utterly disappears when the axion mass is high enough so that $|\Delta_{\|}-\Delta_a|$ is always above $|\Delta_{\mathrm{pl}}|$. Nevertheless, if the plasma density is enhanced with respect to the GJ model, plasma-mass resonance may reappear. In case the plasma effect surpasses the vacuum polarization at the beginning, plasma-mass resonance will be the only resonance point. This scenario may arise in instances involving rapidly rotating neutron stars with a dense plasma, or low-energy photons with weak vacuum polarization. If no resonance condition is met, the conversion is still governed by a non-resonant process.

\section{Polarization Signal and Results}

We now discuss the linear polarization arising from photon-axion mixing. The polarization of photon can be described by the Stokes parameters:
\begin{equation}
\begin{aligned}
I(z) & \equiv  \braket{A_{\perp}(z) A_{\perp}^{*}(z)}+ \braket{A_{\|}(z) A_{\|}^{*}(z)}, \\
Q(z) & \equiv \braket{A_{\perp}(z) A_{\perp}^{*}(z)}- \braket{A_{\|}(z) A_{\|}^{*}(z)}, \\
U(z) & \equiv \braket{A_{\|}(z) A_{\perp}^{*}(z)}+ \braket{A_{\perp}(z) A_{\|}^{*}(z)}, \\
V(z) & \equiv i \left( \braket{A_{\|}(z) A_{\perp}^{*}(z)}-\braket{A_{\perp}(z) A_{\|}^{*}(z)}\right),
\end{aligned}
\end{equation}
and the degree of linear polarization is defined by 
\begin{equation}
\Pi_{L}(z)=\frac{\sqrt{Q^{2}(z)+U^{2}(z)}}{I(z)},
\end{equation}
where the brakets $\braket{\cdots}$ denote the time-averages quantities. Note that the linear polarization parameters $Q$ and $U$ also depend on the choice of phase coordinate system, and we only consider the coordinate system of $U=0$ in this work.

An unpolarized photon originating from the surface of a neutron star undergoes non-resonant photon-axion conversion as it traverses the weak mixing region in the magnetosphere, with the potential for additional resonant conversion at the sphere where the resonance condition is satisfied. This leads to the linear polarization
\begin{equation}
\begin{aligned}
\Pi_L(z) &= \frac{I_{\perp}(z) -(1-P^{\mathrm{res}}_{\gamma \rightarrow a})I_{\|}^{\mathrm{nonres}}(z)}{I_{\perp}(z) +(1-P^{\mathrm{res}}_{\gamma \rightarrow a})I_{\|}^{\mathrm{nonres}}(z)}\\
& \approx \frac{P^{\mathrm{res}}_{\gamma \rightarrow a}+2 |\mathrm{Re}(A_{\|,2})|}{2- (P^{\mathrm{res}}_{\gamma \rightarrow a}+2 |\mathrm{Re}(A_{\|,2})| )},
\end{aligned}
\end{equation}
where $P_{\gamma \rightarrow a }^{\mathrm{res}}$ is the probability of axion-photon resonant conversion. 
The term $|\mathrm{Re}(A_{\|,2})|$ represents the attenuation of the amplitude of $A_{\|}$ from photon-axion non-resonant conversion, in which we consider only the contribution from the leading nontrivial order of the mixing term $\Delta_{\mathrm{M}}$, as shown in the supplementary materials. In our analysis, the conversion probability is always much less than 1.

From the observational side, neutron stars have been subject to excellent measurements at X-ray energies, along with comprehensive multi-wavelength follow-up observations in the radio, optical, and infrared (IR) bands. Here, we examine the optical and IR polarization signals from various neutron stars.

Despite the detection of an increasing amount of neutron stars in this band, the measurements of optical polarization are limited and the IR measurement is still missing. We use the optical polarization of two neutron stars--PSR B0656+14 and 4U 0142+61, as well as the projected IR polarization of a benchmark neutron star. Relevant parameters are detailed in Table~\ref{tab:data}.

\begin{table}[ht]
\begin{tabular}{cccccc}
\hline
Star     & $T$ (s)  & \thead{$r_{\mathrm{NS}}$ \\ (km)} &  \thead{$B_p$\\ ($10^{13}$G)}  & $\omega$ or $\lambda$& PD (\%)       \\ \hline
B0656+14 & 0.384  & 9.3 & 0.47  & 555 nm& 11.9 $\pm$ 5.5~\cite{Mignani:2015ufa}  \\
4U 0142+61& 8.689                    &16.1 %\footnotemark[4]                                 
& 13                              & I-band                                          & $<5.6$~\cite{Wang:2015ppa}     \\
Benchmark IR & 5.0    & 10  & 10 & 0.1 $\mathrm{eV}$  & $<5.0$  \\
 
 \hline
\end{tabular}
\caption{
The key parameters for neutron stars used in this work, including the spin period $T$, radius $r_{\mathrm{NS}}$, magnetic field strength at the surface $B_p$, and the linear polarization degree (PD) measured at the specific energy $\omega$ or wavelength $\lambda$.} 
\label{tab:data}
\end{table}

{\bf PSR B0656+14:} PSR B0656+14 is the fifth brightest optical pulsar to date. The $11.9\% \pm 5.5 \%$ phase-averaged linear optical polarization signal is firstly reported by the Focal Reducer and low dispersion Spectrograph with the high efficiency $v_{\mathrm{HIGH}}$ filter ($\lambda = 555.0 $ nm, $\Delta \lambda = 61.8 $ nm) at Very Large Telescope~\cite{Mignani:2015ufa}. The optical light of B0656+14 is believed to originate from the surface~\cite{Shannon:2004jp,Harding:2017ypk}.

{\bf 4U 0142+61:} The magnetar 4U 0142+61~\cite{2001ApJ...562..918C} has been extensively studied at optical and infrared (IR) wavelengths. An upper limit of 5.6\% at the 90\% confidence level (C.L.) on linear polarization was reported in the optical I-band ($\lambda =802$ nm) using the 8.2-m Subaru Telescope during December 22-23, 2013~\cite{Wang:2015ppa}. The optical I-band emission from 4U 0142+61 may have various origins. However, the observed low linear polarization(at most a few percent) is likely attributable to thermal surface emission~\cite{1988ApJ, Rueda:2013laa, Wang:2015ppa}. We also discuss the possibility of photon emission from an altitude above the neutron star surface in the supplementary material.

{\bf Benchmark IR:}Neutron stars are observed to have thermal emission in X-rays and ultraviolet, indicating surface temperatures from hundreds of thousands to millions of Kelvin. As neutron stars cool down, they may eventually reach the temperature of $\mathcal{O}(100)$~K~\cite{Acevedo:2019agu}. With the advent of advanced infrared telescopes, particularly the JWST, a substantial number of neutron stars exhibiting thermal emission in the infrared band are expected to be discovered in the near future~\cite{Raj:2024kjq}. In the absence of current observation, we take a benchmark neutron star with 5\% linear polarization photon in the Mid-IR band as listed in Table.~\ref{tab:data}.

With the aforementioned formalism at our disposal, we can compute the average degree of linear polarization induced by photon-axion conversion through both non-resonant and resonant processes in their magnetosphere across the phase angle $\Omega t$ and set conservative limits on the axion-photon coupling $g_{a \gamma}$. The constraints are derived by assuming that the axion-induced polarization does not exceed the observed linear polarization, denoted as $a\% \pm b\%$, at the 90\% CL, i.e., $\Pi_{L}^{\mathrm{axion }}(\infty) \leq a
\% + 1.3 b\%  $.  Although there exist astrophysical contributions to the linear polarization, they vary independently from photon-axion conversion over the phase and large cancellation between them is impossible~\cite{Dessert:2022yqq}. 
Moreover, in the absence of angular and directional information for B0656+14 and Benchmark IR, we presume that an observation angle of $\theta_{\mathrm{NS}} = 40^{\circ}$ and the magnetization angle $\theta_{m} = 30^{\circ}$, consistent with the X-ray measurement of 4U 0142+61.

\begin{figure*}[ht]
  \centering
\includegraphics[height=6cm,width=17.5cm]{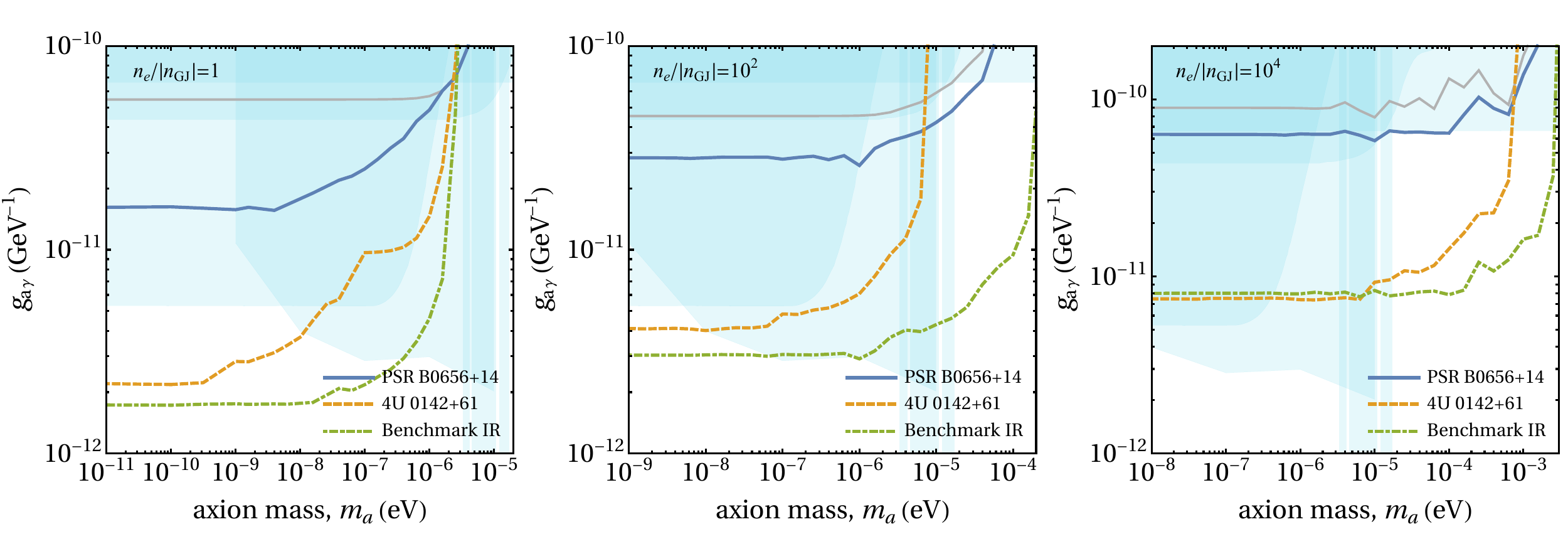}
\caption{ 90\% CL constraints and projected sensitivity on axion-photon coupling $g_{a \gamma}$ from the linear polarization of PSR B0656+14 (blue lines), 4U 0142+61 (orange dashed lines), and Benchmark IR (green dotdashed lines) with different electron densities $n_e = |n_{\mathrm{GJ}}|$ (left panel), $n_e = 10^2 |n_{\mathrm{GJ}}|$ (middle panel), and $n_e = 10^4 |n_{\mathrm{GJ}}|$ (right panel), respectively.  The gray and blue solid lines represent the results of only non-resonant conversion and non-resonant plus resonant conversions for PSR B0656+14, respectively. Shaded regions depict existing constraints from axion-induced X-rays~\cite{Dessert:2021bkv} and polarization~\cite{Dessert:2022yqq} in magnetic white dwarfs, axions produced in pulsar polar-cap cascades~\cite{Noordhuis:2022ljw}, the CAST helioscope~\cite{CAST:2017uph}, ADMX~\cite{ADMX:2018gho, ADMX:2019uok}, and RBF+UF~\cite{DePanfilis:1987dk,Hagmann:1990tj}.}
\label{fig:optical_results}
\end{figure*}

The fiducial limits on $g_{a \gamma}$ for PSR B0656+14 is shown in Fig.~\ref{fig:optical_results}. The contribution of only non-resonant conversion is negligible in most of the parameter space, except for scenarios involving heavier axion masses and low plasma densities so that no resonance point exists, as illustrated in Fig.~\ref{fig:resonance_example}. As the origin of the optical light of 4U 0142+61 is uncertain, we also show the projected sensitivity of 4U 0142+61 and Benchmark IR in dash.

We now focus on the resonance effect. For higher plasma density (from the left to right in Fig.~\ref{fig:optical_results}), the range of axion masses constrained by the polarization signal becomes broader, but the constraints on $g_{a \gamma}$ weaken. This aligns with Fig.~\ref{fig:resonance_example} and the discussion in Sec.~\ref{sec:resonantconversion}, i.e. a strong plasma effect brings the plasma-vacuum resonance point closer to the surface of the neutron star while pushing the plasma-mass resonance away from the star surface, both leading to weaker conversions. For similar reasons, the height of the resonance point $z_{\rm res}$ is higher for 4U 0142+61 than PSR B0656+14, resulting in much stronger constraints for light axions (e.g. $g_{a\gamma} \lesssim 2 \times 10^{-12} \;\mathrm{GeV}^{-1}$ vs $m_a \lesssim 10^{-11}$ eV in the leftmost panel of Fig.~\ref{fig:optical_results}). At high axion masses, $m_a\gtrsim 10^{-5}$~eV, strong limits can be achieved for a more realistic neutron star magnetosphere where the plasma density is higher than that of the GJ model. 

The strength of the resonant conversion depends on the configuration of the magnetosphere, and in particular, the photon energy. Explicitly, Eq.~\eqref{eq:resvacuum} shows $\gamma_{\mathrm{res}} \propto g_{a \gamma}^2 \omega^{-1/3}$, i.e. the adiabaticity ratio at the first resonance point decreases as energy increases. Consequently, there is at most weak resonance contribution in the X-ray band. In the optical bands, 4U 0142+61 potentially imposes stringent constraints on $g_{a \gamma}$ due to its appropriate plasma effects, specifically $g_{a \gamma} \lesssim 3 \times 10^{-12} \;\mathrm{GeV}^{-1}$ for axion masses $m_a \lesssim 10^{-8}$ eV at $n_e = |n_{\mathrm{GJ}}|$, and it could establish new constraints in the axion mass region of $10^{-5}$-$10^{-3}$ eV at $n_e = 10^{4} |n_{\mathrm{GJ}}|$. Moreover, the polarization signals at lower frequencies, such as the mid-IR bands, have greater potential than optical bands to provide a stronger constraint on $g_{a \gamma}$ and to probe a broader range of axion masses. Additional discussions and results are available in the supplementary materials.

\section{Conclusion and discussion}

Although it is generally believed that the excessively strong magnetic field may suppress the axion-photon mixing via the vacuum polarization effect, we demonstrate that strong mixing arising from resonant photon-axion conversion can still take place in the neutron star magnetosphere when the plasma is taken into consideration. By studying neutron star signals in different bands, we find the resonance effect always dominates the conversion whenever present. Lower photon energy and plasma density impose stronger restrictions on $g_{a \gamma}$ in the lower axion mass region, while a strong plasma effect allows to study heavier axions. The polarization signal of neutron stars opens a new window to the exploration of axion-photon coupling near an magnitude below the current limit, without relying on axion as cosmological dark matter.

\section*{Acknowledgments}
We thank Yi-Zhong Fan, Shan-Shan Weng, and Qiang Yuan for their helpful suggestions and discussions. NS is supported by the National Natural Science Foundation of China (NNSFC) No. 12475110, No. 12347105,  No. 12441504 and No. 12047503. LW is supported by the National Natural Science Foundation of China (NNSFC)  No. 12275134, No. 12335005 and No. 12147228.

\bibliography{refs}

%%%%%%%%%% Supplemental materials %%%%%%%%%%
\clearpage
\newpage
\onecolumngrid
\setcounter{page}{1}
\setcounter{equation}{0}
\setcounter{figure}{0}
\setcounter{table}{0}
\setcounter{section}{0}
\setcounter{subsection}{0}
\renewcommand{\theequation}{S.\arabic{equation}}
\renewcommand{\thefigure}{S\arabic{figure}}
\renewcommand{\thetable}{S\arabic{table}}
\renewcommand{\thesection}{\Roman{section}}
\renewcommand{\thesubsection}{\Alph{subsection}}
\newcommand{\ssection}[1]{
    \addtocounter{section}{1}
    \section{\thesection.~~~#1}
    \addtocounter{section}{-1}
    \refstepcounter{section}
}
\newcommand{\ssubsection}[1]{
    \addtocounter{subsection}{1}
    \subsection{\thesubsection.~~~#1}
    \addtocounter{subsection}{-1}
    \refstepcounter{subsection}
}
\newcommand{\fakeaffil}[2]{$^{#1}$\textit{#2}\\}

\thispagestyle{empty}
\begin{center}
    \begin{spacing}{1.2}
        \textbf{\large 
            \hypertarget{sm}{Supplemental Material:} \\ 
            Polarization Signals from Axion-Photon Resonant Conversion in Neutron Star Magnetosphere
        }
    \end{spacing}
    \par\smallskip
    Ningqiang Song$^{1}$,
    Liangliang Su$^{2}$,
    and Lei Wu$^{2}$
    \par
    {\small
        \fakeaffil{1}{Institute of Theoretical Physics, Chinese Academy of Sciences, Beijing, 100190, China}
        \fakeaffil{2}{Department of Physics and Institute of Theoretical Physics, Nanjing Normal University, Nanjing, 210023, China} 

        %(Dated: \today)
    }

\end{center}
\par\smallskip
This supplementary material provides additional details on the polarization calculation and includes further discussions. In Section~\ref{sec:s1}, we present the magnetic field and charge particle density distribution in the neutron star magnetosphere. Section~\ref{sec:s2} details the photon's evolution and conversion in the magnetosphere, covering weak (non-resonant) and strong (resonant) photon-axion mixing regions. In the section.~\ref{sec:s3}, we discuss the constraint on the photon-axion coupling $g_{a \gamma}$ from the polarization signal on X-ray band. Additionally, the impact of photon emission altitude on constraints for the photon-axion coupling parameter $g_{a \gamma}$ is explored in the final section.

\section{Neutron Star Model}\label{sec:s1}
At a distance from the surface of compact star, the magnetic field surrounding a compact star can be commonly described by the dipole magnetic field profile as the higher-harmonic contributions can be neglected, that is
\begin{equation}
\mathbf{B}(\mathbf{r})=\frac{B_{p}}{2}\left(\frac{r_{\mathrm{NS}}}{r}\right)^{3}[3 \hat{\mathbf{r}}(\hat{\mathbf{m}} \cdot \hat{\mathbf{r}})-\hat{\mathbf{m}}],
\end{equation}
where $B_p$ is the magnetic field strength at the surface of star. $\hat{\mathbf{m}}$ denotes the magnetic north pole axis of neutron star. For estimating the plasma effect, in this work, we consider a simplistic charge density model in the magnetosphere of a neutron star, proposed by Goldreich and Julian (GJ)~\cite{goldreich1969pulsar}, 
\begin{equation}
n_{\mathrm{GJ}}\left(\mathbf{r}\right)=\frac{2 \boldsymbol{\Omega} \cdot \mathbf{B}}{e} \frac{1}{1-\Omega^{2} r^{2} \sin ^{2} \theta_{\mathrm{NS}}} .
\label{eq:GJ_model}
\end{equation}
where $ \boldsymbol{\Omega} = \Omega \hat{\mathbf{z}}_{\mathrm{NS}}$ with $\Omega = 2 \pi /T $ and spin period $T$ is the rotation vector of the neutron star, and $\theta_{\mathrm{NS}}$ is the polar angle of observation point with respect to the rotation axis. Meanwhile, if the charge density is dominated by free electrons, i.e., $n_e \approx |n_{\mathrm{GJ}}| $, the plasma frequency can be given by 
\begin{equation}
\begin{aligned}
\omega_{\mathrm{pl}}\left(\mathbf{r}\right) & \approx \sqrt{4 \pi \alpha n_e /m_e}\\
&=\sqrt{B_{z_{\mathrm{NS}}} \frac{4 e \pi}{m_{e} T} \frac{1}{1-\Omega^{2} r^{2} \sin ^{2} \theta_{\mathrm{NS}}}},
\end{aligned}
\end{equation}
where $B_{z_{\mathrm{NS}}}$ is the component of the magnetic field along the $\hat{\mathbf{z}}_{\mathrm{NS}}$ direction, i.e., 
\begin{equation}
    B_{z_{\mathrm{NS}}} = \frac{B_{p}}{2}\left(\frac{r_{\mathrm{NS}}}{r}\right)^{3}\left[3 \cos \theta_{\mathrm{NS}} \hat{\mathbf{m}} \cdot \hat{\mathbf{r}}-\cos \theta_{m}\right],
\end{equation}
where $\theta_m$ is the polar angle of the magnetic axis with respect to the rotation axis. The angle between the magnetic axis $\hat{\mathbf{m}}$ and $\hat{\mathbf{r}}$ is given by, 
\begin{equation}
\hat{\mathbf{m}} \cdot \hat{\mathbf{r}}=\cos \theta_{m} \cos \theta_{\mathrm{NS}}+\sin \theta_{m} \sin \theta_{\mathrm{NS}} \cos (\Omega t),
\end{equation}
where $\Omega t$ also denotes the phase angle of observation.

\section{Photon-Axion Conversion in Neutron Star
Magnetosphere}\label{sec:s2}
In this section, we will describe the conversion of an unpolarized photon from the surface of a neutron star ($z=0$) as it crosses the magnetosphere. The vector potential of an unpolarized photon can be written as 
\begin{equation}
    \mathbf{A}(0) = \frac{\mathcal{A}}{\sqrt{2}}(a_{\|} \hat{\mathbf{x}}_{\|} + a_{\perp} \hat{\mathbf{x}}_{\perp}),
\end{equation}
where $|\mathcal{A}|^2$ denotes the magnitude of the photon. The $a_{\|}$ and $a_{\perp}$ include the phase information of photon, and they need to obey the conditions $\braket{a_{\|} a_{\|}^{*}} = \braket{a_{\perp} a_{\perp}^{*}} = 1 $ and $\braket{a_{\|} a_{\|}} =\braket{a_{\|} a_{\perp}} = \braket{a_{\perp} a_{\perp}} = \braket{a_{\|} a_{\perp}^{*}} = 0$ to ensure $Q=U=V=0$.

\subsection{The Evolution of Photon with Weak Mixing: Non-resonant Conversion}
In the weak mixing regime, the evolution of photon can be obtained by the perturbative solution of the equations of motion (Eq.~\eqref{eq:motion}), where the $\Delta_{\rm M}$ term treated as perturbation, that is 
\begin{equation}
-i \frac{\mathrm{d}}{\mathrm{d} z} \mathbf{\Psi}(z)=(\mathcal{H}_0+\mathcal{H}_{\mathrm{I}}) \mathbf{\Psi}(z),
\label{eq:motion_2}
\end{equation}
here we only consider the evolution of axion and the component of photons parallel to the magnetic field, i.e.,  $\mathbf{\Psi} = (a, A_{\|})^T$ and the Hamiltonians are written as
\begin{equation}
\begin{array}{l}
\mathcal{H}_{0}(z)=\left(\begin{array}{cc}
\Delta_{a} & 0 \\
0 & \Delta_{\|}(z) + \Delta_{\mathrm{pl}}
\end{array}\right) = \Delta_{a} \mathbf{I} + \left(\begin{array}{cc}
0 & 0 \\
0 & \Delta_{\mathrm{tr}} 
\end{array}\right), \\
\\
\mathcal{H}_{\mathrm{I}}(z)=\left(\begin{array}{cc}
0 & \Delta_{\mathrm{M}}(z) \\
\Delta_{\mathrm{M}}(z) & 0
\end{array}\right), 
\end{array}
\end{equation}
where $\Delta_{\mathrm{tr}} = \Delta_{\|}(z) + \Delta_{\mathrm{pl}} - \Delta_a$ and $\mathbf{I}$ is the units matrix. In the interaction representation, the equations of motion can be rewritten as
\begin{equation}
-i \frac{\mathrm{d}}{\mathrm{d} z} \mathbf{\Psi}_{\mathrm{int}}(z)= \mathcal{H}_{\mathrm{int}} \mathbf{\Psi}_{\mathrm{int}}(z),
\end{equation}
where $\mathbf{\Psi}_{\mathrm{int}} = \mathcal{U}^{\dagger} \mathbf{\Psi}$ and $ \mathcal{H}_{\mathrm{int}} = \mathcal{U}^{\dagger} H_{\mathrm{I}} \mathcal{U}$
with 
\begin{equation}
\mathcal{U}(z)=\exp \left[i \int_{0}^{z} \mathcal{H}_{0}\left(z^{\prime}\right) d z^{\prime}\right].
\end{equation}
Therefore, the solution of Eq.~\eqref{eq:motion} can be obtained by iteration:
\begin{equation}
\mathbf{\Psi}_{\mathrm{int}}^{(n+1)}(z)=i \int_{0}^{z} \mathrm{d} z^{\prime} \mathcal{H}_{\mathrm{int}}\left(z^{\prime}\right) \mathbf{\Psi}_{\mathrm{int}}^{n}\left(z^{\prime}\right)
\end{equation}
with the zeroth-order solution $\mathbf{\Psi}_{\mathrm{int}}^{(0)}(z) = \mathcal{U}^{\dagger}(z) \mathbf{\Psi}^{(0)}(z) = (0, A_{\|}(0))^{T}$. Then we have the first-order solution,
\begin{equation}
\begin{aligned}
A_{\|,\mathrm{int}}^{(1)}(z) & = i \int_{0}^{z} \mathrm{d} z^{\prime} \Delta_{\mathrm{M}} e^{-i \int_{0}^{z^{\prime}} \mathrm{d} z^{\prime \prime} \Delta_{\mathrm{tr}}} a_{\mathrm{int}}^{(0)}\left(z^{\prime}\right) = 0, \\ 
a_{\mathrm{int}}^{(1)}(z) & = i A_{\|}(0) \int_{0}^{z} \mathrm{d} z^{\prime} \Delta_{\mathrm{M}} e^{i \int_{0}^{z^{\prime}} \mathrm{d} z^{\prime \prime} \Delta_{\mathrm{tr}}},
\end{aligned}
\label{eq:first-order}
\end{equation}
although the first-order solution can provide the axion-photon conversion rate $p(\gamma \rightarrow a) = |a_{\mathrm{int}}^{(1)}(z)|^2$, we must solve the second-order solution to estimate the state of photon due to $A_{\|,\mathrm{int}}^{(1)}(z)=0$. Then, we have 
\begin{equation}
\begin{aligned}
A_{\|,\mathrm{int}}^{(2)}(z) & = i \int_{0}^{z} \mathrm{d} z^{\prime} \Delta_{\mathrm{M}} e^{-i \int_{0}^{z^{\prime}} \mathrm{d} z^{\prime \prime} \Delta_{\mathrm{tr}}} a_{\mathrm{int}}^{(1)}\left(z^{\prime}\right) \\
& = - A_{\|}(0) \int_{0}^{z} \mathrm{d} z^{\prime} \Delta_{\mathrm{M}}(z^{\prime}) \int_{0}^{z^{\prime}} \mathrm{d} z^{\prime \prime} \Delta_{\mathrm{M}}(z^{\prime \prime})  \exp (-i \int_{z^{\prime \prime}}^{z^{\prime}} \mathrm{d} z^{\prime \prime \prime} \Delta_{\mathrm{tr}}(z^{\prime \prime \prime})).
\end{aligned}
\end{equation}
Finally, removing some unimportant phases, the vector potential of photon after passes through the  distance z in the magnetosphere, is given by 
\begin{equation}
\begin{aligned}
    \mathbf{A}(z) & = \frac{\mathcal{A}}{\sqrt{2}}\left[a_{\perp} \hat{\mathbf{x}}_{\perp}+ a_{\|} \hat{\mathbf{x}}_{\|} \left(1- A_{\|,2}(z)\right) \right]\\
\end{aligned}
\label{eq:Photon_Az}
\end{equation}
with 
\begin{equation}
    A_{\|,2}(z) = \int_{0}^{z} \mathrm{d} z^{\prime} \Delta_{\mathrm{M}} (z^{\prime})  \int_{0}^{z^{\prime}} \mathrm{d} z^{\prime \prime} \Delta_{\mathrm{M}}(z^{\prime \prime}) e^{-i \int_{z^{\prime \prime}}^{z^{\prime}} \mathrm{d} z^{\prime \prime \prime} \Delta_{\mathrm{tr}}(z^{\prime \prime \prime})}.
\end{equation}
The Eq~.\eqref{eq:first-order} and ~\eqref{eq:Photon_Az} are not only applicable to scenario of the inhomogeneous magnetic filed, but also can obtain the form of the homogeneous magnetic filed in the previous studies. 

\subsection{The Evolution of Photon with Strong Mixing: Resonant Conversion}
However, these non-resonant conversion results will break down in strong mixing regime, particularly the resonance region~\cite{PhysRevD.37.1237}. Now, we consider the axion-photon resonance in the inhomogeneous magnetic filed. The eigenvalues and eigenvectors of equation of motion (Eq.~\eqref{eq:motion_2}) can be defined by
\begin{equation}
\begin{aligned}
k_{ \pm}&=\frac{\Delta_{\|}+\Delta_{a}+\Delta_{\mathrm{pl}}}{2} \pm \frac{\Delta k}{2},\\
\Delta k &= \left[\left(\Delta_{a}-\Delta_{\|}-\Delta_{\mathrm{pl}}\right)^{2}+ 4\Delta_{M}^{2}\right]^{1 / 2}, \\
\left(\begin{array}{c}
a \\
A_{\|}
\end{array}\right)_{+}&=\left(\begin{array}{c}
\cos \beta_{m} \\
\sin \beta_{m}
\end{array}\right),\left(\begin{array}{c}
a \\
A_{\|}
\end{array}\right)_{-}=\left(\begin{array}{c}
-\sin \beta_{m} \\
\cos \beta_{m}
\end{array}\right).
\end{aligned}
\end{equation}
For the axion-photon resonance, $\Delta_a = \Delta_{\|} +\Delta_{\mathrm{pl}}$, there is $\beta_m =\pm 45^{\circ}$ and the mode evolution depends on the adiabaticity ratio $\gamma = \Delta_k /|2 \beta_m^{\prime}|$, where the $\beta_m^{\prime}$ is given by 
\begin{equation}
\beta_m^{\prime} = \left(\frac{1}{2 \tan 2 \beta_m} +\frac{2 \Delta_{\|} +\Delta_{\mathrm{pl}} }{4 \Delta_{\mathrm{M}}}\right) \frac{1}{H} \sin^2 2 \beta_m  
\end{equation}
with the density scale height along the l.o.s $H = \rho_e/ |\rho_e^{\prime}| \approx B/|B^{\prime}|$. Thus, at the resonance point, the adiabaticity ratio is 
\begin{equation}
\gamma_{\mathrm{res}} = \frac{4 \Delta_M^2 H}{|\Delta_a+\Delta_{\|}|}. 
\end{equation}
Given that the nonadiabaticity exclusively manifests at the axion-photon resonance point, the Landau-Zener formula can be used to calculate the nonadiabatic jump probability, 
\begin{equation}
P_{\mathrm {jump }}=e^{-\pi \gamma_{\text {res }} / 2} .
\end{equation}
The probability of axion-photon conversion across the resonance is expressed as $P_{\gamma \rightarrow a} = 1 - P_{\mathrm{jump}}$ and is positively correlated with the resonance adiabaticity ratio $\gamma_{\mathrm{res}}$.

\section{X-ray Polarization Signal}\label{sec:s3}

\begin{figure*}[ht]
  \centering
\includegraphics[height=7cm,width=14cm]{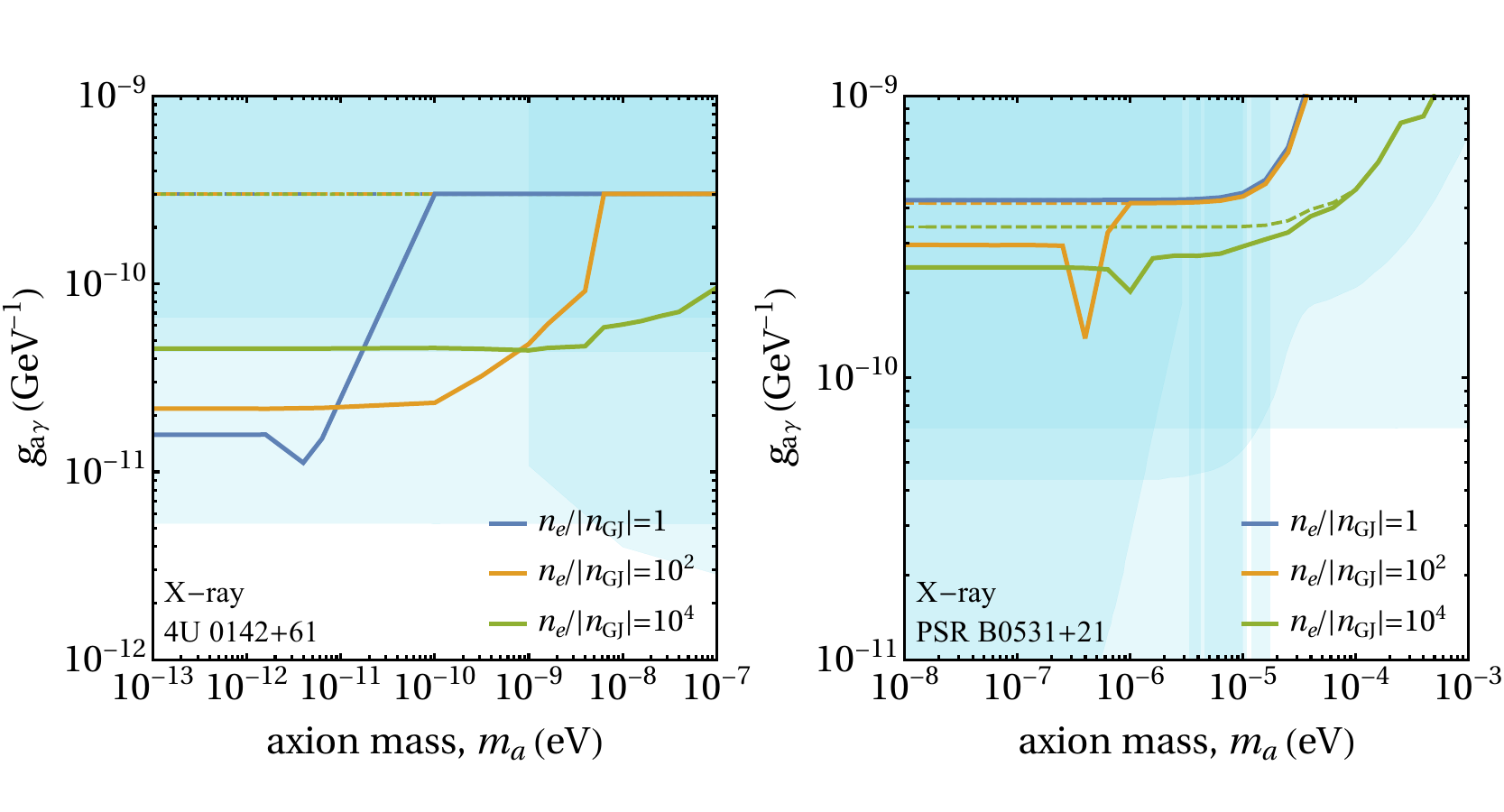}
\caption{As in Fig.~\ref{fig:optical_results} but for X-ray linear polarization signals for the PSR B0531+21 (right plane) and 4U 0142+61 (left plane ) at different electron densities $n_e = |n_{\mathrm{GJ}}|$ (blue lines), $n_e = 10^2 |n_{\mathrm{GJ}}|$ (orange lines), and $n_e = 10^4 |n_{\mathrm{GJ}}|$ (green lines).  The dashed and solid lines represent the results of only non-resonant conversion and non-resonance plus resonant conversion,
respectively. }
\label{fig:xray_results}
\end{figure*}

As early as the 1970s, the linear polarization of X-ray bands from the Crab has been measured using the Bragg polarimeter onboard OSO-8~\cite{1976ApJ...208L.125W,1978ApJ...220L.117W}. In our study, we utilize the recently reported polarization results obtained by a miniature X-ray polarimeter onboard the PolarLight, which employs a new high-sensitivity technique~\cite{Feng:2020miz}. The PolarLight was launched on October 29, 2018, and the measurements indicate an average linear polarization of approximately $15.3 \pm 3\%$ for photons within the energy range of $3-4.5$ keV. While this result encompasses information from Crab nebula emission, we consider it valid for our calculations as the analysis results for pulsar + nebula and pulsar-free nebula emission are consistent within the margin of error. Simultaneously, we consider another X-ray linear polarization signal from the magnetar 4U 0142+61. The Imaging X-ray Polarimetry Explorer (IXPE) has recently provided the first-ever measurement of polarized emission from 4U 0142+61 in the X-ray bands~\cite{Taverna:2022jgl}. The results indicate a linear polarization degree of $14 \pm 1\%$ across the IXPE 2-4 keV energy range, which comes from the directions of $\theta_{\mathrm{NS}} = 40^{\circ} $ and $\theta_{m} = 30^{\circ}$ obtained by the rotating vector model.

Applying the same analysis method as used for the optical bands, we illustrate the results of X-ray bands polarization signals for 4U 0142+61 (left plane) and the Crab pulsar (right plane) at different electron densities $n_e = |n_{\mathrm{GJ}}|$ (blue lines), $n_e = 10^2 |n_{\mathrm{GJ}}|$ (orange lines), and $n_e = 10^4 |n_{\mathrm{GJ}}|$ (green lines) in Fig.~\ref{fig:xray_results}.  For 4U 0142+61, the behavior of results in X-ray bands is similar to that of optical bands, except the constraints on $g_{a\gamma}$ and the upper limit of axion mass become weaker. This arises from the adiabaticity ratio of the first resonance point (major contribution) having the relation: $\gamma_{\mathrm{res}} \propto g_{a \gamma}^2 \frac{r_{\mathrm{NS}}+z_{\mathrm{res}}}{ \omega} \approx g_{a \gamma}^2 \omega^{-1/3}$ for $\Delta_{\|} = |\Delta_{\mathrm{pl}}|$ with $\omega \propto (r_{\mathrm{NS}}+z_{\mathrm{res}})^{3/2}$. This implies that higher energy photons require a stronger coupling to achieve the same conversion probability, leading to weaker constraints. Moreover, the maximum axion mass that can be constrained is also lower for X-ray observations than the optical band. This is because as the axion mass increases, the two types of resonance points become close to each other and eventually no resonance is encountered. For higher energy photons, the plasma-vacuum resonance point occurs further away from the surface, corresponding to lower plasma density. As a result, the maximum axion mass becomes smaller ($m_a^2 \approx \omega_{\mathrm{pl}}^2$).

In the X-ray band, for the rapidly rotating Crab pulsar, the resonance point $r_{\mathrm{NS}}+z_{\mathrm{res}} \propto \omega_{\mathrm{pl}}(0)^{-2/3} \omega^{2/3}$ is highly likely to occur outside the magnetosphere (light cylinder  radius $R_{\mathrm{LC}} = 1/\Omega = P/(2 \pi)$) for lower plasma density. Thus, in this case, there is only the contribution of non-resonant conversion (blue line), and as the plasma density increases, the resonant conversion begins to take effect, as shown in the right plane of Fig.~\ref{fig:xray_results}.

\section{The influence of emission altitude on polarization signals}

\begin{figure*}[ht]
  \centering
\includegraphics[height=6cm,width=17.5cm]{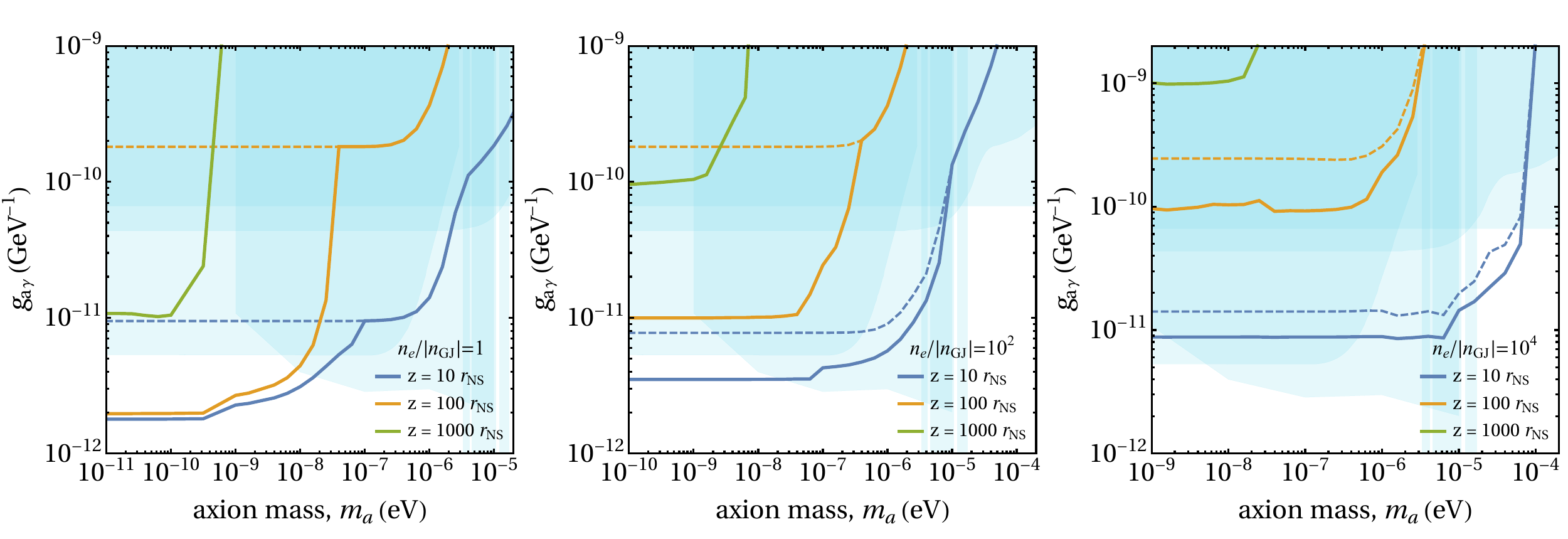}
\caption{The 90\% C.L. sensitivity on axion-photon mixing ($g_{a \gamma}$) from optical band polarization signals of 4U 0142+61. Results are presented for various electron densities: $n_e = |n_{\mathrm{GJ}}|$ (left), $n_e = 10^2 |n_{\mathrm{GJ}}|$ (middle), and $n_e = 10^4 |n_{\mathrm{GJ}}|$ (right). The analysis assumes the photon originates from heights $z = 10\;r_{\mathrm{NS}}$ (blue lines), $z = 100\; r_{\mathrm{NS}}$ (orange lines), and $z = 1000\; r_{\mathrm{NS}}$ (green lines) above the neutron star surface.}
\label{fig:height_results}
\end{figure*}

Compared to the X-ray emission of neutron stars, the origin of optical and other band emissions is complex and not well understood. Some studies suggest that the optical emission includes a thermal spectral component from the surface of neutron star and a nonthermal component from its magnetosphere~\cite{Zharikov:2021llh}. Here, we take the optical polarization signal of 4U 0142+61 as an example to discuss the impact of emission altitude on the constraints of $g_{a \gamma}$. It is worth noting that the optical emission from this source is likely to originate from the surface due to its low degree of polarization (at most a few percent)~\cite{Wang:2015ppa}. Fig.~\ref{fig:height_results} suggests that the contribution of non-resonant conversion will decrease sharply with increasing emission altitude due to the weakening of the strength of the magnetic field. In resonant conversion, the resonance points are closer to the surface of the neutron star for the heavier axions and larger plasma densities, as shown in Fig.~\ref{fig:resonance_example}. Therefore, assuming a photon with an emission altitude $z = 1000 \;r_{\mathrm{NS}}$, the axion-photon coupling is constrained to $g_{a\gamma} \lesssim 10^{-11}$ eV at the electron density $n_e = |n_{\mathrm{GJ}}|$ for $m_a \lesssim 10^{-10}$ eV, but it weakens to $g_{a \gamma} \lesssim 10^{-9}$ at $n_e = 10^4 |n_{\mathrm{GJ}}|$ for $m_a \lesssim 10^{-8}$ eV. And the results will weaken as the emission altitude increases until the emission altitude exceeds all possible resonance points. Meanwhile, in rapidly rotating neutron stars like the Crab pulsar, the influence of emission altitude is more significant compared to slowly rotating neutron stars, primarily because the larger plasma density make the resonance points closer to the neutron star surface, as indicated by Eq.~(\eqref{eq:GJ_model}).
\end{document}